\documentclass[aps,prl,twocolumn,groupedaddress,showpacs,reprint]{revtex4-1}
\usepackage{graphicx}
\usepackage{amsmath}
\usepackage{amssymb}
\usepackage{color}

\usepackage[normalem]{ulem}

\newcommand{\exciting}{{\usefont{T1}{lmtt}{b}{n}exciting }}
\newcommand{\excitingn}{{\usefont{T1}{lmtt}{b}{n}exciting}}

\begin{document}

\title{Microhartree Precision in Density-Functional-Theory Calculations}

\author{Andris Gulans$^1$, Anton Kozhevnikov$^2$, and Claudia Draxl$^1$}
\affiliation{$^1$Physics Department and IRIS Adlershof, Humboldt-Universit\"at zu Berlin, Zum Gro\ss en Windkanal 6, 12489 Berlin, Germany}
\affiliation{$^2$Swiss National Supercomputing Center, Lugano, Switzerland}

\date{\today}

\begin{abstract}
To address ultimate precision in density-functional-theory calculations we employ the full-potential 
linearized augmented planewave + local-orbital (LAPW+lo) method and justify its usage as a benchmark method.
LAPW+lo and two completely unrelated numerical approaches, multi-resolution analysis 
(MRA) and linear combination of atomic orbitals, yield total energies of atoms with a mean
deviation of 0.9~$\mu$Ha and 0.2~$\mu$Ha, respectively. 
Spectacular agreement with the MRA is reached also for total and atomization energies of the G2-1 set consisting of 55 molecules. 
With the example of $\alpha$-iron we demonstrate the capability of LAPW+lo of reaching $\mu$Ha/atom 
precision also for periodic systems, which allows also for distinction between numerical precision and the accuracy of a given functional.
\end{abstract}



\maketitle

Density-functional theory (DFT)~\cite{Hohenberg64,Kohn1965a} is the most widely used method in modern computational condensed-matter physics and chemistry, 
as reflected by the existence of dozens of implementations, employing diverse numerical schemes. 
While each of these implementations may be most suitable for a specific type of applications, in essence, 
all of them solve the Kohn-Sham (KS) equation~\cite{Kohn1965a}. 
Given the diversity of computational tools, it is natural to ask whether they indeed provide the same answers. 
This question, i.e., how reproducible DFT results are, was recently discussed in the context of a community effort, 
where the equation of state of 71 elemental solids was calculated using a variety of DFT 
implementations~\cite{Lejaeghereaad3000}. 
It turned out that, after a period of fine-tuning, different DFT codes are now in good overall agreement. 
Despite the deviations between codes being small on average, discrepancies obtained for certain elements are 
much more substantial. Moreover, it is not clear, or even not expected that such agreement would be preserved beyond 
bulk elemental materials. As a matter of fact, the work to be done to explore and guarantee the precision of 
electronic-structure codes is certainly far from being finished.

Efforts on the improvement of computational methods are immensely alleviated if reliable reference data or 
reference tools are available. The gold standard for solving the same KS equation of DFT for condensed matter 
are full-potential all-electron methods, especially those employing (linearized) augmented planewaves with local 
orbitals (LAPW+lo)~\cite{Slater1937,Andersen1975,Sjoestedt2000} as basis functions.
LAPW+lo is often \textit{trusted blindly} as the ultimate reference method for validating pseudopotentials or data 
sets of projector-augmented waves~\cite{Hamann2013,Garrity2014,Jollet2014,Schlipf2015}. 
Yet, it has never been shown how precise this method can be in practice. Even more, arguing that 
the method would depend on parameters which "can influence the results in a more or less erratic way",
it was even questioned recently ~\cite{Jensen2017} whether LAPW+lo can provide benchmark quality at all.

In this Letter, we use our LAPW+lo implementation in the full-potential all-electron package 
\exciting \cite{Gulans2014a} to show that for absolute total energies outstanding
1~$\mu$Ha/atom precision can be obtained. In order to validate this statement, 
we first turn to atoms and molecules, since for finite systems one can find other methods that, in 
principle, are capable of yielding an {\it exact} numerical solution of the KS equation. 
In the second step, we exploit the duality of the LAPW+lo basis for verifying the numerical 
performance of planewaves using atomic orbitals and vice versa. Making use of this concept, we demonstrate that 
$\mu$Ha precision is achievable also for periodic systems. 
Clearly, we can thus distinguish between the accuracy of a DFT functional and the numerical precision of the actual implementation.

Let us recall the Kohn-Sham equation of DFT, 
\begin{equation}
 \left[-\frac{\nabla^2}{2}+v_\mathrm{eff}(\mathbf{r})\right]\psi(\mathbf{r})=\varepsilon_\mathrm{KS}\psi(\mathbf{r}).
\end{equation}
The major source of numerical issues in solving it, is the behavior of the effective potential 
$v_\mathrm{eff}(\mathbf{r})$. While it is very smooth in most of the space, its shape is dominated 
by the electrostatic contribution in the proximity of nuclei, where it varies rapidly with a 
divergence at the nuclear sites. As a result, the otherwise well-behaved KS orbitals 
$\psi(\mathbf{r})$ exhibit cusps at the atomic positions and a nodal structure in their vicinity. 

The LAPW+lo method meets these properties of $\psi(\mathbf{r})$. 
The unit cell is partitioned into non-overlapping atomic spheres (or muffin-tin spheres, \textit{MT$_{\alpha}$}), 
centered at the nuclear positions, with index $\alpha$, and the interstitial region (\textit{I}).
KS wavefunctions are expanded in terms of atom-like functions, 
$f^\alpha_\nu(\mathbf{r_\alpha})=u_\nu(r_\alpha)Y_{\ell m}(\hat{\mathbf{r}}_\alpha)$, 
and planewaves, respectively:
\begin{equation}
\label{eq:lapw}
\phi_\mathbf{G+k}(\mathbf{r})= \left\{
\begin{array}{cl}
  \sum\limits_{\nu} A^\alpha_{\mathbf{G+k},\nu} f^\alpha_\nu(\mathbf{r_\alpha}) & \mathbf{r}\in MT_\alpha\\
  \frac{1}{\sqrt{\Omega}} e^{i(\mathbf{G+k})\mathbf{r}} & \mathbf{r}\in I
\end{array}
\right. .
\end{equation}
The coefficients $A^\alpha_{\mathbf{G+k},\nu}$ are determined to ensure smoothness of the basis functions at the sphere boundaries. 
These augmented planewaves are typically complemented by local orbitals:
\begin{equation}
\label{eq:lo}
\phi_\mu(\mathbf{r})= \left\{
\begin{array}{cl}
  f^\alpha_\mu(\mathbf{r}) & \mathbf{r}\in MT_\alpha\\
  0 & \mathbf{r}\in I
\end{array}
\right. .
\end{equation}
That way the flexibility of the basis is improved which, indeed, has a major impact on results as we demonstrate below. 
Local orbitals are crucial also for reaching benchmark quality in $GW$ calculations as was pointed out recently \cite{Friedrich2011,Nabok2016}.
A more detailed introduction to the LAPW+lo method is available in Refs.~\cite{Singh1994,Gulans2014a}. 
The overall size of the basis and the quality in the interstitial region are controlled by the 
dimensionless parameter $R_\mathrm{MT}G_\mathrm{max}$, where $G_\mathrm{max}$ is the maximum length of wavevectors $\mathbf{G+k}$ used in the LAPW basis.
In other words, $R_\mathrm{MT}G_\mathrm{max}$ can be freely adjusted to make the expansion of 
wavefunctions in the interstitial region as precise as necessary.
In the atomic spheres, the quality of the wavefunctions is governed by the choice of the 
atomic-like functions (Eqs. \ref{eq:lapw} and \ref{eq:lo}).

\begin{figure}
 \includegraphics[width=\columnwidth]{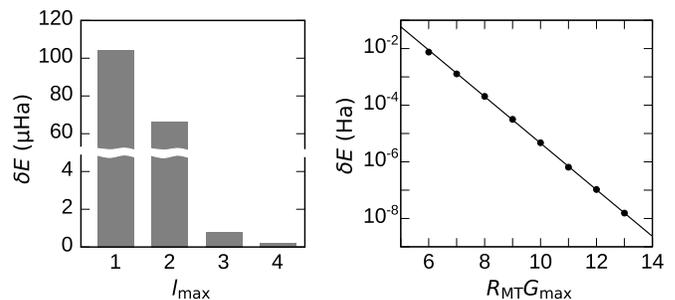}
 \caption{\label{fig:lo-lmax} Error in the total LSDA energy of an oxygen atom when using local orbitals 
 with angular momenta up to $\ell_\mathrm{max}$ (left) and as a function of a planewave cutoff 
 $R_\mathrm{MT}G_\mathrm{max}$ (right). The limit of the total energy is estimated by using 
 $\ell_\mathrm{max}=6$ and by extrapolating its dependence on $R_\mathrm{MT}G_\mathrm{max}$.
} 
 \end{figure}

To illustrate how the LAPW+lo basis can be exploited to reach essentially exact total energies for a given 
exchange-correlation functional, we consider the oxygen atom. We restrict ourselves to using the local 
spin-density approximation (LSDA). Still, the same procedure works for any other functional, for which a local KS 
potential is available, and we present a similar discussion for the generalized-gradient approximation (GGA) in the 
Supplemental Material~\cite{SM}. According to the Aufbau principle, the $2p$-shell is partially filled, with one $p$-
orbital doubly and two others singly occupied. Consequently, this atom is not only magnetic, but its effective 
potential and hence the electron density are not spherically symmetric. Thus, radial and angular degrees of freedom are 
entangled, and all wavefunctions formally contain contributions from all angular momenta. 
We take it into account by introducing local orbitals not only with angular momenta of $\ell$=0 and 1, that are the 
dominating contributions to the $1s$, $2s$, and $2p$ states, but consider also higher values of $\ell$. 
Their impact on the total energy is shown in Fig.~\ref{fig:lo-lmax}. 
In the classical LAPW formalism, $f^\alpha_\nu(\mathbf{r})$ combines strictly two functions per spherical harmonic 
for each atom. 
Local orbitals allow us to correct for all missing features in the pure LAPW representation and are not limited in number.
A calculation using local orbitals with $\ell$ up to 1 yields the total energy within an error of $\sim 100~\mu$Ha. 
Adding basis functions with higher angular momenta gradually improves the result, and, at $\ell_\mathrm{max}=4$, this quantity differs from the estimated exact limit by less than $10^{-7}$~Ha. 
Likewise, we show how the total energy converges with increasing $R_\mathrm{MT}G_\mathrm{max}$. 
In this case, we observe an exponential decay of the error similarly as in Ref.~\cite{Gulans2014a}. 
Note that the errors due to missing angular degrees of freedom depend on the atomic-sphere radius; 
the magnitudes shown in Fig.~\ref{fig:lo-lmax} are obtained for $R_\mathrm{MT}=1.2~a_0$. 
It is possible to reduce $R_\mathrm{MT}$ so far that already at $\ell_\mathrm{max}=1$ the errors are negligible.
We find that, at $R_\mathrm{MT}=0.5~a_0$, this error is only $2~\mu$Ha.
However, a reduction of $R_\mathrm{MT}$ to such a small value leads to an enormous increase of the number of LAPWs, making the calculations very expensive.

At this point, it is already clear that the quality of the basis in the atomic spheres is essential 
for highly accurate results, and the discussed example reflects how to handle systems with 
substantially asymmetric potentials in the atomic spheres. Such potentials not only occur in a 
range of spin-polarized systems, but also in systems with short bonds.

To show that the converged limit in LAPW+lo corresponds to the exact numerical solution we compare 
them to two other all-electron methods that are expected to deliver highly precise results.
The first one is the multi-resolution analysis (MRA)~\cite{Harrison2004}.
It recasts the KS equation in the Lippmann-Schwinger integral equation form and solves it 
iteratively by applying local and non-local operators on trial wavefunctions numerous times. 
The wavefunctions are stored in an adaptive multi-scale representation, while the integral kernels 
of the non-local operators are represented in a separable form. 
Such a numerical approach allows for solving the KS equation efficiently with a guaranteed precision. 
The MRA is implemented in the MADNESS code~\cite{Harrison2004}, which is currently restricted to finite systems.
The other alternative method is the linear combination of atomic orbitals (LCAO), for which we use the NWChem package~\cite{Valiev2010}. 
Although, in the general case, the precision of LCAO for absolute total energies is limited, 
it was shown how Gaussian-type orbitals can be used for reaching the complete-basis limit for atoms~\cite{Schmidt1979}
which we employ in this work.
In the calculations of molecules, we resort to the augmented correlation-consistent polarized 
quadruple- and quintuple-$\zeta$ basis sets known also by their acronyms \texttt{aug-cc-pVQZ} and 
\texttt{aug-cc-pV5Z} ~\cite{Kendall1992,Woon1993,Prascher2011}, respectively. These basis sets were used by 
Willand \textit{et al.} \cite{Willand2013} for generating all-electron reference data for 
benchmarking newly-generated pseudopotentials. The so-obtained atomization energies turned out to 
be converged to at least 1~kcal/mol ($\approx 1.6\cdot 10^{-3}$~Ha), which is commonly 
referred to as chemical accuracy. 

\begin{figure}
\includegraphics[width=\columnwidth]{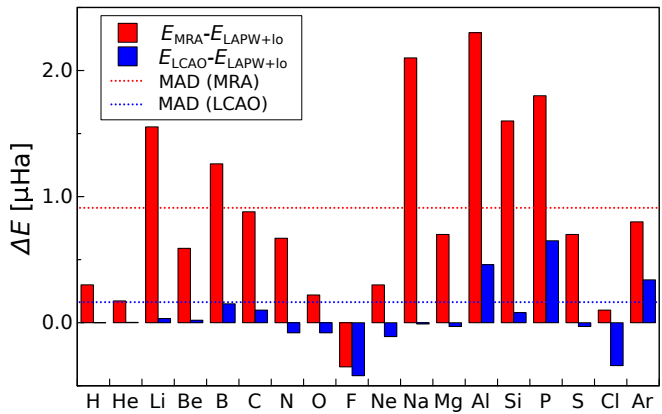}
\caption{\label{fig:atoms} (Color online) 
The LSDA energies of atoms calculated using MADNESS (MRA) and NWChem (LCAO). 
The \exciting (LAPW+lo) results are used as the reference.
Note the $\mu$Ha precision throughout.
} 
 \end{figure}

MRA and LCAO are designed for calculating finite systems, and thus we make use 
of them first for comparison with our total energies of atoms and further below for molecules. 
We employ non-relativistic theory and the LSDA~\cite{Perdew1992}. 
This choice, however, does not influence the overall conclusions from our work.
The total energies of atoms obtained with the three codes are compared in Fig.~\ref{fig:atoms}.
%from the first three periods are listed in Table~\ref{tab:atoms}. 
In all cases, we observe outstanding agreement between LAPW+lo and MRA  (LCAO) with a 
mean absolute deviation of 0.9~$\mu$Ha (0.2~$\mu$Ha). 
(Computational details and the total energies are provided in the Supplemental Material \cite{SM}.)

The excellent agreement between the different methods is all the more remarkable in view of the very different ways 
the KS equation is solved. In particular, it concerns the fact that \exciting has been developed primarily for 
studying problems of condensed-matter theory, e. g., it considers electrons in periodic potentials.
In other words, we compute isolated atoms and molecules employing periodic boundary conditions. 
It translates into a huge number of basis functions, and, thus, requires an efficient way to diagonalize the Hamiltonian.
To do so, we have implemented a novel approach. It follows the block-Davidson algorithm~\cite{Davidson1975}, though
with an important modification such that the initial subspace is particularly suitable for LAPW+lo.
It consists of an initial guess for the KS wavefunctions, all local orbitals and a number of eigenvectors of the 
overlap matrix. As a result, we obtain an algorithm that predictably converges even at high values of the cutoff 
parameter $R_\mathrm{MT}G_\mathrm{max}$ when the LAPWs become nearly linearly dependent. Our implementation of 
this algorithm follows the spirit of Ref.~\cite{Goedecker1992} and thus does not require the construction of the entire 
Hamiltonian and overlap matrices. Details of the implementation will be published elsewhere \cite{Gulans2017}.

In order to reach the limit of an isolated molecule (or atom) in LAPW+lo calculations, it is necessary to ensure that there is no artificial interaction between the periodic images of molecules in neighboring unit cells. 
It is especially critical for polarizable molecules with permanent dipoles, as their interaction energy scales as $d^{-3}$~\cite{Makov1995}, where $d$ is the distance between adjacent molecules. 
We eliminate this slow decay by truncating the Coulomb potential~\cite{Rozzi2006}. 
Such an approach is particularly important for molecules like LiH.
The truncation of the Coulomb potential allows for a reduction of the unit-cell dimensions from $80$~\AA{} to $16$~\AA.
Thus, the size of the LAPW+lo basis reduces from $10^8$ to $10^6$ making the total-energy calculation feasible.
Note that even $10^6$ basis functions is an unusually large problem size in comparison to typical LAPW+lo calculations.

Equipped with this methodology, we turn to the second benchmark, which is the G2-1 set~\cite{Curtiss1997a}. 
This set contains 55 molecules, consisting of 2--8 atoms for which a variety of experimental data is available. 
Thus it provides an excellent opportunity for benchmarking methods of DFT and quantum chemistry. 
Here, we use it for comparing different computational methods. 
We consider fixed geometries according to the data published in Refs.~\cite{G2geom,Curtiss1997}.

\begin{table}
 \caption{\label{tab:G2}
Mean deviation (MD), mean absolute deviation (MAD), and maximum absolute deviation (MaxD) of LAPW+lo total 
(left columns) and atomization energies (right columns) of the G2-1 molecules with respect to the results obtained with MADNESS (MRA) and NWChem (LCAO).
All quantities are obtained using the LSDA and given in Ha/atom.}
 \begin{center}   \setlength{\tabcolsep}{0.2cm}
 \begin{tabular}{l|cc|cc}
 \hline
    & \multicolumn{2}{c|}{Total energy} & \multicolumn{2}{c}{Atomization energy} \\
    & $\Delta E^{tot}_\mathrm{MRA}$ & $\Delta E^{tot}_\mathrm{LCAO}$ & $\Delta E^{at}_\mathrm{MRA}$ & $\Delta E^{at}_\mathrm{LCAO}$ \\
  \hline
MD   & $0.2\cdot 10^{-6}$ & $1.2\cdot 10^{-3}$ & $0.4\cdot 10^{-6}$ & $6.1\cdot 10^{-5}$ \\
MAD  & $0.3\cdot 10^{-6}$ & $1.2\cdot 10^{-3}$ & $0.5\cdot 10^{-6}$ & $9.9\cdot 10^{-5}$ \\
MaxD & $1.1\cdot 10^{-6}$ & $7.6\cdot 10^{-3}$ & $1.5\cdot 10^{-6}$ & $2.1\cdot 10^{-3}$ \\
  \hline
 \end{tabular}
 \end{center}
\end{table}

Table~\ref{tab:G2} summarizes deviations of the MRA and LCAO energies from those obtained with \excitingn.
The complete list of total energies can be found in the Suplemental Material \cite{SM}. 
The agreement between LAPW+lo and MRA is spectacular for both absolute total energies and atomization energies.
The average and maximum deviations in total energy from the other methods amount to 
0.3~$\mu$Ha/atom and 1.1~$\mu$Ha/atom, respectively, consistent 
with the results for atoms shown above. Similarly, we obtain 0.5~$\mu$Ha/atom and 
1.5~$\mu$Ha/atom for the average and maximum deviations in atomization energies, 
respectively. The excellent agreement between the highly converged LAPW+lo and MRA calculations
allows us to argue that these two methods provide essentially exact answers. The obtained 
discrepancy, thus, can be considered as the error estimate of the two methods. 

\begin{figure}
\includegraphics[width=0.9\columnwidth]{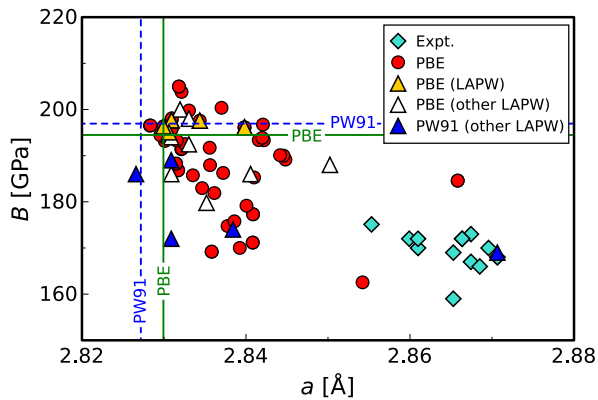}
\caption{\label{fig:iron} (Color online) Bulk moduli and lattice constants of $\alpha$-iron.
Yellow triangles correspond LAPW+lo calculations, red circles to results by other methods taken from Ref. \cite{Lejaeghereaad3000}. 
Empty and blue triangles represent older LAPW(+lo) calculations~\cite{Zhang2010}. Lines indicate results obtained in this work. 
Diamonds correspond to experimental data~\cite{Zhang2010,Rayne1961,Rotter1966,Leese1968,Simmons1971,Dever1972,Jephcoat1986,Ghosh2002,Klotz2000,Adams2006,Speich1972,Acet1994,Lejaeghere2014}.
} 
 \end{figure}

As argued above, the employed LCAO basis sets are not supposed to yield exact answers. 
Indeed, the average absolute deviations in the total (atomization) energies of LAPW+lo from LCAO calculations 
are three (two) orders of magnitude larger compared to those from the MRA. Still, with the only 
exception of the SO$_2$ molecule (error of $1.9\cdot 10^{-3}~\mu$Ha (1.2 kcal/mol)), the LCAO 
atomization energies are within the chemical-accuracy limit of 1.0~kcal/mol. 

The high precision of the LAPW+lo method obtained in calculations of atoms and molecules suggests a similar performance for solids. 
While it would be desirable to benchmark against other methods also for periodic systems, we are not aware of implemented alternative approaches that are expected to be exact. 
On the other hand, the nature of the LAPW+lo method opens a door for ``self-validation'' through the partitioning of space that introduces two very different ways of representing wavefunctions. 
More specifically, if the atomic-sphere volume is reduced, a certain region that was previously described by atomic-like orbitals is now described by planewaves. 
We argue that if such a rearrangement does not introduce a change in the total energy, the complete-basis limit has been reached.

We demonstrate the performance of the LAPW+lo method for periodic systems with the example of ferromagnetic $\alpha$-iron. 
This material presents numerical challenges as reflected in Ref.~\onlinecite{Lejaeghereaad3000}, 
where the corresponding results show scattering well above the average over the 71 elements of the Periodic Table.
The need for a precise and thus unique result given a certain functional is also motivated by the fact that all-electron calculations are commonly taken as a reference for benchmarking pseudopotentials,
as it was done in a recent DFT study of thermoelastic properties of iron~\cite{Dragoni2015}.

The aspherical density distribution due to the magnetic moment, as discussed above, 
requires particular care. In order to handle the anisotropy of the potential in the atomic spheres we introduce 
local orbitals with angular momenta $\ell$ up to 6. Using this setup and a sufficiently high LAPW cutoff, 
we vary $R_\mathrm{MT}$ in the range of 1.4--2.1~$a_0$. Note that such a variation of the atomic sphere corresponds 
to a change in its volume by a factor of three. Remarkably, the total energy stays within the 1~$\mu$Ha range. 
%(see Table \ref{tab:iron}). 
This is the case for any considered volume within 10\% deviation from the equilibrium 
volume of the primitive unit cell. Thus, we argue that the 1~$\mu$Ha precision has been achieved also for this case.
The outstanding agreement is obtained for the entire energy-versus-volume dependence, 
as demonstrated by the Birch-Murnaghan equation of state based on 21 data points within $\pm 5\%$ of volume change, shown in in Table~\ref{tab:iron}. 
As soon as convergence is reached in terms of the planewave cutoff, we obtain extremely stable values 
of the equilibrium volume $V_0$, the bulk modulus $B_0$, and its pressure derivative $B^\prime$. The former two fluctuate 
only in their sixth, the latter one in its fifth decimal place. A polynomial fit considering a wider range of 
volumes ($\pm 10\%$) exhibits the same stability (see Supplemental Information \cite{SM}). 

\begin{table}
 \caption{\label{tab:iron}
Equilibrium volume, $V_0$ (in atomic units), bulk modulus, $B_0$ (in GPa), and its pressure derivative, 
$B^\prime$, for $\alpha$-iron as obtained from a fit of non-relativistic LSDA results to the Birch-Murnaghan equation of state. $\Delta E^{tot}$ (in $\mu$Ha) is relative to the value in the first row. 
All results above the separating line are fully converged.}
 \begin{center} \setlength{\tabcolsep}{0.15cm}  
 \begin{tabular}{cccccr}
 \hline
 $R_{MT}$ & $R_\mathrm{MT}G_\mathrm{max}$ & $V_0$ & $B_0$ & $B^\prime$ &  $\Delta E^{tot}$\\
 \hline                                                                   
1.4 & 14 & 71.3298 & 236.296 & 4.5992 & 0.0 \\
%1.5 & 14 & 71.3299 & 236.297 & 4.5996 & 0.1 \\
%1.6 & 14 & 71.3299 & 236.295 & 4.5998 & 0.2 \\
%1.7 & 14 & 71.3299 & 236.294 & 4.5994 & 0.3 \\
1.8 & 14 & 71.3299 & 236.295 & 4.5994 & 0.4 \\
%1.9 & 14 & 71.3300 & 236.295 & 4.5989 & 0.5 \\
%2.0 & 14 & 71.3301 & 236.296 & 4.5989 & 0.6 \\
2.1 & 14 & 71.3302 & 236.295 & 4.5994 & 0.8 \\
\hline
%2.1 & 13 & 71.3301 & 236.293 & 4.5986 & 1.2 \\
2.1 & 12 & 71.3297 & 236.298 & 4.5998 & 3.8 \\
%2.1 & 11 & 71.3285 & 236.311 & 4.5965 & 18.1 \\
2.1 & 10 & 71.3243 & 236.337 & 4.5968 & 95.0 \\
%2.1 & 9 & 71.2907 & 236.157 & 4.5855 & 490.1 \\
2.1 & 8 & 71.1087 & 239.443 & 4.6320 & 2253.0 \\
\hline
%  \hdashline
 \end{tabular}
 \end{center}
\end{table}

Using the same settings as above, we obtain also highly precise values for the equilibrium lattice constant and bulk modulus from scalar-relativistic PBE calculations, shown in Fig.~\ref{fig:iron} together with data from the literature.
The scattering of the calculations (wider than the experimental ones) do not allow for conclusions about the {\it exact} result unless a highly reliable reference calculation is available. 
Strikingly, our reference value obtained in this work is located far from the middle of the \textit{cloud} of the PBE data from Refs.~\cite{Lejaeghereaad3000} and \cite{Zhang2010}, and farthest away from experiment. 
Notably, our results allow for comparing performance of two functionals as we illustrate by discussing PBE and PW91~\cite{Perdew1991}.
The spread of data obtained with the two GGAs implies that, without reference data, a distinction between the accuracy of a given functional and the numerical precision introduced by a specific implementation would not be possible. 
Indeed, besides a few exceptions~\cite{Mattsson2006}, these two GGAs were often considered synonymous. 
Our calculations clearly show that PW91 yields a smaller lattice constant and a larger bulk modulus than PBE.

In conclusion, we have challenged the numerical accuracy of the LAPW+lo method. In order to demonstrate its capability, 
calculations for atoms and molecules have been benchmarked against two completely unrelated, highly precise methods. 
The differences in absolute total energies are on average $1~\mu$Ha/atom. Furthermore, we have shown that 
we reach the same precision also for solids. The presented results allow us to claim that, once properly converged, 
LAPW+lo is an essentially exact method for DFT calculations. Overall, this work presents also a justification for 
using LAPW+lo as a reference method, backing up its reputation as the \textit{gold-standard} method of DFT for condensed matter. 
The ability to reach the complete-basis limit will be indispensable for benchmarking less precise methods and for quality control of data collections. 
Furthermore, it opens perspectives towards reliably computing numerically sensitive quantities, like magnetisation anisotropy, weak non-covalent interactions, relative stabilities of isomers or polymorphs etc., where high precision is crucial.

\begin{acknowledgments}
The work has received partial support from the European Union's Horizon 2020 research and innovation programme, 
grant agreement No. 676580 through the Center of Excellence NOMAD (Novel Materials Discovery Laboratory) \cite{NOMAD-CoE}.
A. K. acknowledges helpful discussions with Robert J. Harrison on the MRA method.
\end{acknowledgments}


\begin{thebibliography}{49}
\bibitem [{\citenamefont {Gulans}()}]{Gulans2017}%
  \BibitemOpen
  \bibfield  {author} {\bibinfo {author} {\bibfnamefont {A.}~\bibnamefont
  {Gulans}},\ }\href@noop {} {\enquote {\bibinfo {title} {Application of
  {D}avidson diagonalization algorithm to augmented plane waves},}\ }\bibinfo
  {note} {To be published}\BibitemShut {NoStop}%
\end{thebibliography}
\end{document}